\documentclass[preprint,showpacs,preprintnumbers,endfloats*,amsmath,amssymb,prb]{revtex4-1}
\usepackage{graphicx}
\usepackage{dcolumn}
\usepackage{bm}

\begin{document}

\preprint{}

\title{Interplay between O defects and SiC stacking at the SiC/SiO$_2$ interface}

\author{Christopher James Kirkham}
\email{kirkham@ccs.tsukuba.ac.jp}
\affiliation{Center for Computational Sciences, University of Tsukuba, Tsukuba, Ibaraki 305-8577, Japan}
\author{Tomoya Ono}
\email{ono@ccs.tsukuba.ac.jp}
\affiliation{Center for Computational Sciences, University of Tsukuba, Tsukuba, Ibaraki 305-8577, Japan}
\affiliation{JST-PRESTO, Kawaguchi, Saitama 332-0012, Japan}

\date{\today}

\begin{abstract}
We investigate the effect of SiC stacking on the 4H-SiC/SiO$_2$ interface, both in the presence and absence of O defects, which appear during thermal oxidation, via first principles calculations. It is known that 4H-SiC(0001) has two different surface types, depending on which of the two lattice sites, $h$ or $k$, is at the surface [K. Arima \textit{et al}., Appl. Phys. Lett. \textbf{90}, 202106 (2007)]. We find interlayer states along the conduction band edge of SiC, whose location changes depending on the interface type, and thus too the effect of defects. When $h$ sites are directly at the interface, O defects remove interfacial conduction band edge states. On the other hand, when $k$ sites are at the interface, the conduction band edge is insensitive to the presence of O defects. These differences will impact on the operation of SiC devices because the most commonly used SiC based metal-oxide-semiconductor field-effect transistors rely on the electronic structure of the conduction band.
\end{abstract}

\maketitle
Silicon carbide (SiC) is a wide band-gap semiconductor material that can be used for high temperature and power electronic devices, in conditions where the traditionally used Si fails. These improved properties are due to the presence of strong Si-C bonds. Like Si, its native oxide is SiO$_2$, which can be grown via thermal oxidation, making it useful for metal-oxide-semiconductor field-effect transistors (MOSFETs), primarily of the n-channel type. However, its use in practical devices has been hampered by the low channel mobility of the SiC/SiO$_2$ interface compared with bulk SiC~\cite{Afanas'ev1997,Afanas'ev2004}. This has been partly attributed to a variety of C and O defects found on both sides of the interfacial region~\cite{Knaup2005,Deak2007,Gavrikov2008,Ono2015}. Traditionally research has focused on defects which contribute interface gap states, but advances in interface growth techniques mean that these gap states no longer make a significant contribution to reductions in mobility~\cite{Dhar2010,Fiorenza2013,Liu2015}. Further advances require new information.

\begin{figure}
\begin{center}
\includegraphics{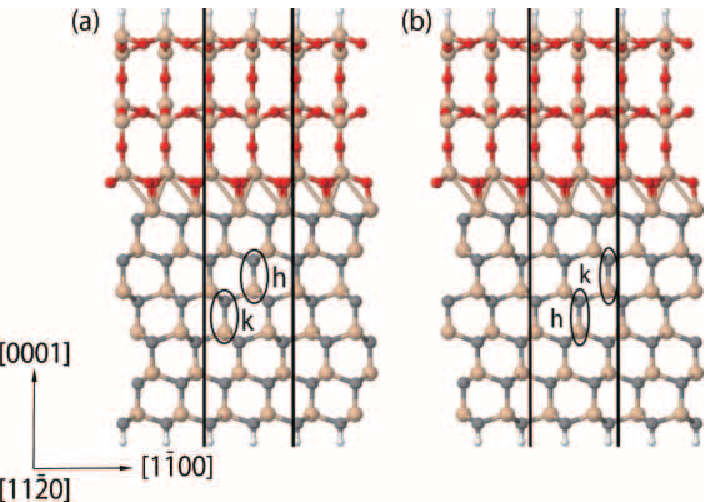}
\caption{(Color online) SiC/SiO$_2$ interface with (a) $h$ and (b) $k$ lattice sites at the interface. Black lines indicate the bounding box of the calculation cell. Highlighted Si-C pairs show examples of $h$ and $k$ sites. Si atoms are beige, C grey, O red and H white.}
\label{fig:SiC}
\end{center}
\end{figure}

SiC has many different polytypes, the most commonly used of which is 4H-SiC. This consists of four repeated SiC bilayers, with two inequivalent lattice sites for the Si and C atoms. Based on their local symmetries, these are often labelled $h$ for hexagonal and $k$ for quasi-cubic, as shown in Fig.~\ref{fig:SiC}. At $h$ sites the connected SiC bilayers have Si-C bonds in opposite directions, whilst for $k$ sites these Si-C bonds run parallel. The local structure around the SiC/SiO$_2$ interface will change depending on which of these sites is at the SiC surface~\cite{Arima2007,Arima2011}. For convenience these will be referred to as $h$ and $k$ type interfaces based on the site which the surface Si atom occupies. Previous work looking at bulk SiC reported interlayer \emph{floating} states at the conduction band edge (CBE)~\cite{Matsushita2012,Matsushita2014}, but have not reported the effect of defects at the SiC/SiO$_2$ interface. Floatings states appear between subsequent SiC bilayers with the same orientation.

Here we investigate how the energies and local density of states (LDOS) of the interface changes with interface type, and when O atoms are subsequently introduced. This is important for real devices, because currently the surface SiC bilayer cannot be controlled during the thermal oxidation process. Due to differences in local structure, floating states are observed at the interface for $h$ type, but from the second bilayer down for $k$ type. More interestingly, O defects behave differently in the two interfaces, removing the floating states at the $h$ type whilst leaving the $k$ type unchanged. The variation of the floating states at the CBE is expected to affect the carrier mobility at the interface because the operation of n-channel MOSFETs relies on the electronic structure at the CBE.

Calculations are carried out using RSPACE~\cite{Hirose2005}, which performs density functional theory~\cite{Hohenberg1964,Kohn1965} calculations using a real-space finite-difference method~\cite{Chelikowsky1994}, employing a time-saving double-grid technique~\cite{Ono1999,Ono2005,Ono2010}. The electron-ion interactions are described using the projector-augmented wave method~\cite{Blochl1994} for the C, Si and O atoms, and a norm-conserving pseudopotential~\cite{Kleinman1982,Troullier1991} for H. The exchange-correlation interaction is treated within the local density approximation~\cite{Vosko1980}.

The interface is modelled using 6 bilayers of 4H-SiC and 9~\AA~thick of $\beta$-tridymite SiO$_2$, with the interface made at the SiC(0001) surface, saturating all the Si dangling bonds with O. Both the top and bottom layers of the model are H terminated. The interface is constructed using experimental values for the Si-C and Si-O bond lengths. The lateral size of the cell is SiC(0001) $3\times3\sqrt{3}$ with the $z$ direction perpendicular to the SiC surface. The cell is periodic in all directions with an 11~\AA~vacuum gap in the $z$ direction. Calculations are performed with a coarse grid spacing of 0.16~\AA~and using the $\Gamma$ and M k-points of SiC(0001) $3\times3\sqrt{3}$. The bottom H atoms and SiC bilayer are held fixed and structural optimization is performed on all remaining atoms to reduce the forces between atoms. A variety of known O defects~\cite{Knaup2005,Deak2007,Gavrikov2008,Ono2015} are introduced into the SiC. Full details on interface atomic structures including defects will be described later.

Formation energies for defects are calculated according to
\begin{equation}
E_{form}=E_{(n-1)\text{O}}+\mu_\text{O}- E_{n\text{O}}
\label{eq:E_form}
\end{equation}
where $E_{n\text{O}}$ is the energy of the interface containing $n$ excess O atoms and $\mu_\text{O}$ is the chemical potential of oxygen, calculated from the energy of an O$_2$ molecule. The energies of single O defects are compared to the clean interface, whilst double O defects are compared to the lowest energy single O defect.

Formation energies can reveal which defect structures are more likely to form, giving an idea of their relative abundance at the interface, but do not give an insight how their properties differ from the clean interface. Thus we also examine the LDOS for both types of interface, with and without defects. The LDOS is calculated according to

\begin{equation}
\rho(z,E)=\sum_{i,k} \int |\Psi_{i,k}(x,y,z)|^2 \mathrm{d}x \mathrm{d}y \times N e^{-\alpha(E-\varepsilon_{i,k})^2}
\label{eq:LDOS}
\end{equation}
where $\varepsilon_{i,k}$ are the eigenvalues of the wavefunction, with indexes of $i$ and $k$ for the eigenstate and the k-point respectively. $N=2\sqrt{\frac{\pi}{\alpha}}$ is the normalisation factor, with $\alpha$ as the smearing factor, here set to 13.5~eV$^{-2}$.

\begin{figure}
\begin{center}
\includegraphics{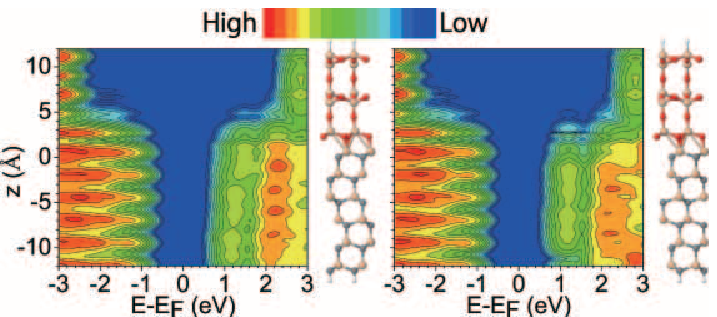}
\caption{(Color online) LDOS for the $h$ (left) and $k$ (right) type interface with no defects. The horizontal axis is energy relative to the Fermi energy, E$_F$, defined as the middle of the band gap. The vertical axis is the height of the model. For clarity, structural models are provided to the right of each LDOS. Contours correspond to density, from blue at the lowest to red at the highest, doubling at each contour line, from a minimum of 5.3$\times 10^{-7}$~states eV$^{-1}$\AA$^{-1}$.}
\label{fig:Clean_LDOS}
\end{center}
\end{figure}

Before looking at defects, it is useful to compare the clean interface. We find that the $h$ type interface is lower in energy than the $k$ type by 0.27~eV per $3\times3\sqrt{3}$ region. The LDOS for these two structures, as shown in Fig.~\ref{fig:Clean_LDOS}, is mostly the same, except for along the CBE, where several oval-like features can be seen, whose location changes with interface type. For $h$ type, one of these states is at the interface, whereas for $k$ type the first state does not appear until the second bilayer. This changes the band gap in the vicinity of the interface. These CBE states are the previously noted floating states~\cite{Matsushita2012,Matsushita2014}, with the difference in location a reflection of the local structure at the interface. To aid with future discussions on this matter, we introduce channel depth, a measure of the number of subsequent SiC bilayers with the same orientation. Channel depth is essentially the same concept as the channel length discussed in Ref.~\onlinecite{Matsushita2014}, but the procedure for counting the length of a channel is different. We use a slightly different term because our counting procedure is more intuitive for describing the interface. For bulk 4H-SiC the longest channel depth is two. At the $h$ type interface the channel depth is also two, but for the $k$ type interface the depth is only one. A minimum channel depth of two is required to observe floating states. This distinction between interface types is important, especially given the position of interfacial defects.

\begin{figure}
\begin{center}
\includegraphics{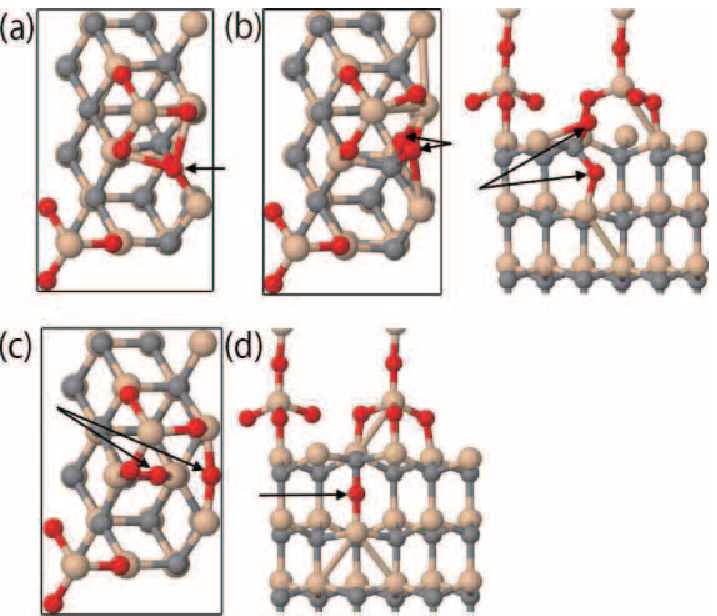} 
\caption{(Color online) Structural models for the various O defects in the $h$ type interface. (a) O$_{if}$ from above. (b) O$_{sub}$+O$_{if}$ from both above and the side. (c) V$_\text{C}$O$_2$ from above. (d) O$_{sub}$ from the side. Views from above look down on the SiC(0001) surface, with the SiO$_2$ above the connecting region removed for clarity. Black boxes indicate the bounding box of the calculation cell. Side views looking down the [1$\bar{1}$00] direction are to clarify the position of subsurface atoms. Arrows indicate defect positions.}
\label{fig:Defects}
\end{center}
\end{figure} 

We then introduce O atoms into the interface as shown in Fig.~\ref{fig:Defects}. In our previous study~\cite{Ono2015}, we investigated the total energies of all possible O atom sites and obtained the most stable configurations. Here, the lowest energy structures which appear during one oxidation cycle are examined. For a single excess O this is O$_{if}$~\cite{Knaup2005} [Fig.~\ref{fig:Defects}(a)], an interstitial site at the interface. For two excess O atoms this is O$_{sub}$+O$_{if}$~\cite{Gavrikov2008} [Fig.~\ref{fig:Defects}(b)], which combines subsurface and interstitial sites. Post CO emission this is V$_\text{C}$O$_2$~\cite{Deak2007} [Fig.~\ref{fig:Defects}(c)], which involves O atoms passivating the dangling bonds created by the C vacancy.

The formation energies of defects are calculated for both interface types, with energy orderings agreeing with prior calculations~\cite{Ono2015}. Formation energies for O$_{if}$ are 0.17 and 0.25~eV for the $h$ and $k$ type interfaces respectively. For O$_{sub}$+O$_{if}$ these formation energies are 0.60 and 0.83~eV. In most cases formation energies are greater in the $k$ type interface, despite the clean interface being higher in energy. This means $k$ type interfaces would have a slightly higher defect density, but as the following LDOS show, this is of little consequence because its electronic structure is largely insensitive to O defects.

\begin{figure}
\begin{center}
\includegraphics{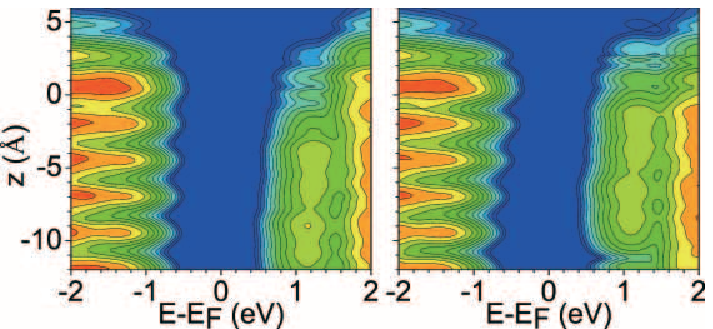}
\caption{(Color online) LDOS for the O$_{if}$ defect in the $h$ (left) and $k$ (right) type interface. Notation is the same as Fig.~\ref{fig:Clean_LDOS}. These LDOS focus on the region around the SiC band gap.}
\label{fig:Single_O_LDOS}
\end{center}
\end{figure}

In addition to their energies, we also examine the LDOS for each defect structure. Firstly O$_{if}$, as shown in Fig.~\ref{fig:Single_O_LDOS}, where the most significant changes occur at the CBE, and only for the $h$ type interface. The first floating state is removed, shifting the position of the CBE and increasing the band gap. The presence of an extra O atom has a channel blocking effect, reducing the channel depth at the interface, and removing the floating state. The $k$ type interface shows no significant changes at the CBE because there is no floating state to remove. It is also useful to study a single subsurface site, O$_{sub}$ [Fig.~\ref{fig:Defects}(d)], to see how O location influences the channel blocking effect. The same changes at the CBE are observed for O$_{sub}$ (not presented here), meaning that defects located within a bilayer or between two bilayers both have a channel blocking effect. On the other hand, changes at the valence band edge (VBE) between the clean interface and O$_{if}$ are the same for both interface types. A defect state is observed at the VBE, but the state is not distinct from the bulk SiC states. This defect state can be attributed to a C--O $\pi$ bond, as observed in the partial DOS for the highest occupied band (not presented here). Because the dissociation energy of a C--O bond at the SiC/SiO$_2$ interface is lower than that of a Si--C bond,\cite{Akiyama2015} the VBE at the interface shifts up in energy slightly.

\begin{figure}
\begin{center}
\includegraphics{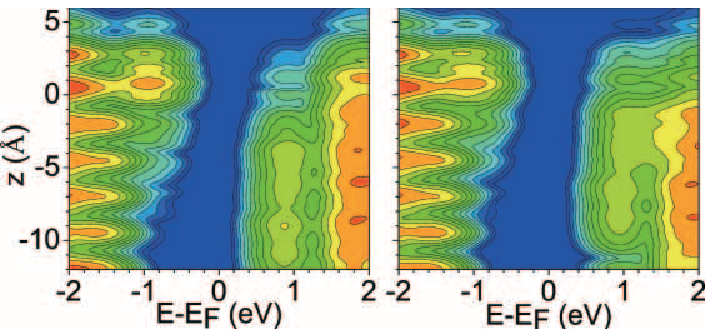}
\caption{(Color online) LDOS for the O$_{sub}$+O$_{if}$ defect in $h$ (left) and $k$ (right) type interface. Notation is the same as Fig.~\ref{fig:Clean_LDOS}. These LDOS focus on the region around the SiC band gap.}
\label{fig:Double_O_LDOS}
\end{center}
\end{figure}

Secondly, O$_{sub}$+O$_{if}$ as shown in Fig.~\ref{fig:Double_O_LDOS}. Changes at the CBE are in line with earlier results, with the interface floating state removed for $h$ type, but $k$ type mostly unaffected. This supports the role of O defects in channel blocking. Changes at the VBE are more noticeable than previous results, with a defect state which is distinct from the bulk SiC. Examining the partial DOS for the highest occupied bands shows that this defect state again arises from C--O $\pi$ bonds. The important difference between O$_{sub}$+O$_{if}$ and O$_{if}$ is that it contains an O--C--O bond, which the latter does not. As such, the states associated with O$_{sub}$+O$_{if}$ are even higher in energy, making them distinct from the VBE.

\begin{figure}
\begin{center}
\includegraphics{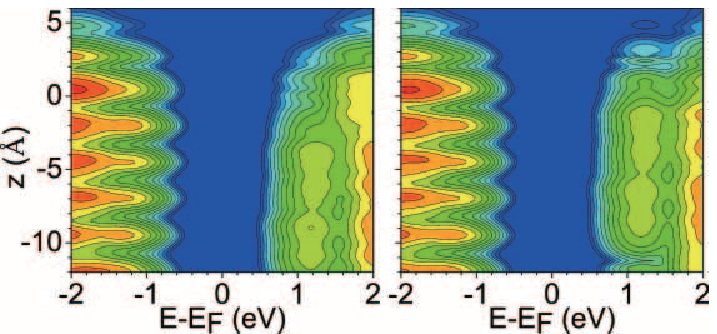}
\caption{(Color online) LDOS for the V$_\text{C}$O$_2$ defect in $h$ (left) and $k$ (right) type interface. Notation is the same as Fig.~\ref{fig:Clean_LDOS}. These LDOS focus on the region around the SiC band gap.}
\label{fig:VCO2_LDOS}
\end{center}
\end{figure}

Finally, the LDOS for the V$_\text{C}$O$_2$ structure, as shown in Fig.~\ref{fig:VCO2_LDOS}, which looks similar to the clean interface. Once again, at the CBE the floating state at the $h$ type interface is removed due to channel blocking by the defect, with no change for the $k$ type. In addition, no changes are observed at the VBE for both interface types. There are no defect states because there are no C--O bonds present. 

We have also examined some of the C-related defect structures reported in Refs.~\onlinecite{Knaup2005} and~\onlinecite{Deak2007} and found defect gap states which appear close to the CBE, partially obscuring changes to the floating states. However, if results for C-related defects are included our overall conclusions do not change.

\begin{figure}
\begin{center}
\includegraphics{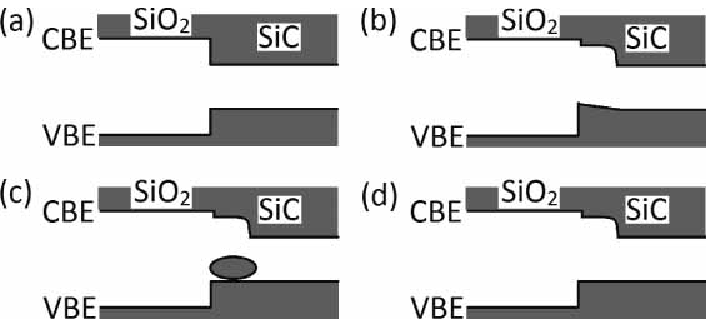}
\caption{Schematic band diagrams for the $h$ type SiC/SiO$_2$ interface, highlighting important changes caused by defects. (a) Clean interface. (b) Single O defect. (c) Double O defect (d) Post CO emission.}
\label{fig:schematic_bands}
\end{center}
\end{figure}

These results are summarized in Fig.~\ref{fig:schematic_bands}, with the primary focus on the $h$ type interface, since it undergoes greater changes in the presence of defects. The clean $h$ type interface is represented by (a). When defects are introduced, the band diagram is modified in two main ways, as shown in (b), (c), and (d) for the O$_{if}$, O$_{sub}$+O$_{if}$, and V$_\text{C}$O$_2$ defects respectively. Firstly, states at the CBE are removed, opening the band gap near the interface. This occurs due to channel blocking by O defects, and occurs in all cases reported here. The gap at the CBE extends over the first two bilayers because the state that would appear there is removed. Secondly a defect state can appear near the VBE, reducing the band gap. For single O atoms, such as O$_{if}$, the band edge at the interface shifts up in energy slightly, as shown in (b). The defect state is only distinct from the bulk SiC states in the case of O$_{sub}$+O$_{if}$, as shown in (c).

A similar gap is observed at the CBE of the $k$ type interface, whether defects are present or not, but with a smaller extent. The gap only extends over the first bilayer, a reflection of the channel depth at the interface. If the extent of this gap is ignored, both the clean interface and the V$_\text{C}$O$_2$ defect can be depicted by Fig.~\ref{fig:schematic_bands}(d), with the O$_{if}$ and O$_{sub}$+O$_{if}$ defects represented by (b) and (c) respectively. However, diagram (a) never appears for the $k$ type interface. 

In summary, we find that interface type determines whether CBE floating states at the interface are affected by the presence of O defects, as with $h$ type, or not, as with $k$ type. Floating states are observed between adjacent SiC bilayers with parallel Si--C bonds and are found from the first bilayer down for the $h$ type interface and from the second bilayer down for the $k$ type. When O defects are introduced at the interface they have a channel blocking effect, removing the floating state at the $h$ type interface, with little effect on the $k$ type interface, where there are no states to remove. Defect states are also found at the VBE, arising from the C--O $\pi$ bonds formed by the O defects, but are only distinct from the bulk SiC states in the case of the O$_{sub}$+O$_{if}$ defect. These states do not vary between the $h$ and $k$ type interfaces. Overall the CBE of the $h$ type interface is more affected by defects, so it would be expected that floating states will influence the electronic properties of n-channel MOSFET, such as carrier mobility. In the future we want to do more complete calculations for all of these structures, to get a fuller picture of how the properties of the interface change with defect type.

This work was supported by the Computational Materials Science Initiative (CMSI). The numerical calculations were carried out using the computer facilities of the Institute for Solid State Physics at the University of Tokyo.


\begin{references}
\bibitem{Afanas'ev1997}
VV. Afanas'ev, M. Bassler, G. Pensl, and M. Schulz, Phys. Status Solidi A {\bf 162}, 321 (1997).

\bibitem{Afanas'ev2004}
VV. Afanas'ev, F. Ciobanu, S. Dimitrijev, G. Pensl, and A. Stesmans, J. Phys.: Condens. Matter {\bf 16}, S1839 (2004).

\bibitem{Knaup2005}
J. Knaup, P. De\'{a}k, T. Frauenheim, A. Gali, and Z. Hajnal, and W. J. Choyke, Phys. Rev. B {\bf 71}, 235321, (2005).

\bibitem{Deak2007}
 P. De\'{a}k,  J. Knaup, T. Hornos, C. Thill, A. Gali, and T. Frauenheim, J. Phys. D {\bf 40}, 6242 (2007).

\bibitem{Gavrikov2008}
A. Gavrikov, A. Knizhnik, A. Safonov,  A. Scherbinin, A. Bagaturfyants, and B. Potapkin, A. Chatterjee, and K. Matocha, J. Appl. Phys. {\bf 104}, 093508 (2008).

\bibitem{Ono2015}
T. Ono and S. Saito, Appl. Phys. Lett. {\bf 106}, 081601 (2015).

\bibitem{Dhar2010}
S. Dhar, S. Haney, L. Cheng,  S.-R. Ryu, A. K. Agarwal, L. C. Yu, and K. P. Cheung, J. Appl. Phys. {\bf 108}, 054509 (2010).

\bibitem{Fiorenza2013}
P. Fiorenza, F. Giannazzo, M. Vivona, A. La Magna, and F. Roccaforte, Appl. Phys. Lett. {\bf 103}, 153508 (2013).

\bibitem{Liu2015}
G. Liu, B.R. Tuttle, and S. Dhar, Appl. Phys. Rev. {\bf 2}, 021307 (2015).

\bibitem{Arima2007}
K. Arima, H. Hara, J. Murata, T. Ishida, R. Okamoto, K. Yagi, Y. Sano, H. Mimura, and K. Yamauchi, Appl. Phys. Lett. {\bf 90}, 202106 (2007).

\bibitem{Arima2011}
K. Arima, K. Endo, K. Yamauchi, K. Hirose, T. Ono, and Y. Sano, J. Phys.: Condens. Matter {\bf 23}, 394202 (2011).

\bibitem{Matsushita2012}
Y. Matsushita, S. Furuya, and A. Oshiyama, Phys. Rev. Lett. {\bf 108}, 246404 (2012).

\bibitem{Matsushita2014}
Y. Matsushita and A. Oshiyama, Phys. Rev. Lett. {\bf 112}, 136403 (2014).

\bibitem{Hirose2005}
K. Hirose, T. Ono, Y. Fujimoto, and S. Tsukamoto, {\it First Principles Calculations in Real-Space Formalism, Electronic Configurations and Transport Properties of Nanostructures} (Imperial College, London, 2005).

\bibitem{Hohenberg1964}
P. Hohenberg and W. Kohn, Phys. Rev. {\bf 136}, B864 (1964).

\bibitem{Kohn1965}
W. Kohn and L. J. Sham, Phys. Rev. {\bf 140}, A1133 (1965).

\bibitem{Chelikowsky1994}
J. R. Chelikowsky, N. Troullier, and Y. Saad, Phys. Rev. Lett. {\bf 72}, 1240 (1994).

\bibitem{Ono1999}
T. Ono and K. Hirose, Phys. Rev. Lett. {\bf 82}, 5016 (1999).

\bibitem{Ono2005}
T. Ono and K. Hirose, Phys. Rev. B {\bf 72}, 085115 (2005).

\bibitem{Ono2010}
T. Ono, M. Heide, N. Atodiresei, P. Baumeister, S. Tsukamoto, and S. Bl\"ugel, Phys. Rev. B {\bf 82}, 205115 (2010).

\bibitem{Blochl1994}
P.E. Bl\"ochl, Phys. Rev. B {\bf 50}, 17953 (1994).

\bibitem{Kleinman1982}
L. Kleinman and D.M. Bylander, Phys. Rev. Lett. {\bf 48}, 1425 (1982).

\bibitem{Troullier1991}
N. Troullier and J.L. Martins, Phys. Rev. B {\bf 43}, 1993 (1991).

\bibitem{Vosko1980}
S. H. Vosko, L. Wilk, and M. Nusair, Can. J. Phys. {\bf 58}, 1200 (1980).

\bibitem{Akiyama2015}
T. Akiyama, A. Ito, K. Nakamura, T. Ito, H. Kageshima, M. Uematsu, K. Shiraishi, Surf. Sci. {\bf 641}, 174 (2015).
\end{references}
\end{document}